\documentclass[sn-mathphys,Numbered]{sn-jnl}

\usepackage{graphicx}%
\usepackage{multirow}%
\usepackage{amsmath,amssymb,amsfonts}%
\usepackage{amsthm}%
\usepackage{mathrsfs}%
\usepackage{caption}
\usepackage[title]{appendix}%
\usepackage{xcolor}%
\usepackage{textcomp}%
\usepackage{manyfoot}%
\usepackage{booktabs}%
\usepackage{algorithm}%
\usepackage{algorithmicx}%
\usepackage{algpseudocode}%
\usepackage{listings}%

\newcommand{\be}{\begin{equation}}
\newcommand{\ee}{\end{equation}}

\begin{document}

\title[Kaniadakis entropy in extreme gravitational and cosmological environments: a review on the state-of-the-art and future prospects]{Kaniadakis entropy in extreme gravitational and cosmological environments: a review on the state-of-the-art and future prospects}


\author*[1]{\fnm{Giuseppe Gaetano} \sur{Luciano}}\email{giuseppegaetano.luciano@udl.cat}

\affil*[1]{\orgdiv{Department of Chemistry, Physics and Environmental and Soil Sciences}, \orgname{Escola Polit\`ecninca Superior, Universidad de Lleida}, \orgaddress{\street{Av. Jaume
II, 69}, \city{Lleida}, \postcode{25001}, \country{Spain}}}


\abstract{Kaniadakis ($\kappa$-deformed) statistics is being widely used for describing relativistic systems with non-extensive behavior and/or interactions. It is built upon a one-parameter generalization of the classical Boltzmann-Gibbs-Shannon entropy, possessing the latter as a particular sub-case. 
Recently, Kaniadakis model has been adapted to accommodate the complexities of systems under the influence of gravity. The ensuing framework exhibits a rich phenomenology that allows for a deeper understanding of the most extreme conditions of the Universe. Here we present the state-of-the-art of $\kappa$-statistics, discussing its virtues and drawbacks at different energy scales.
Special focus is dedicated to gravitational and cosmological implications, including effects on the expanding Universe in dark energy scenarios. This review highlights the versatility of Kaniadakis paradigm, demonstrating its broad application across various fields and setting the stage for further advancements in statistical modeling and theoretical physics. 
}




\maketitle

\section{Introduction}
Boltzmann-Gibbs (BG) theory forms the bedrock of classical statistical mechanics, providing a robust framework to connect the microscopic properties of multi-degree-of-freedom systems to their observable behaviors. Nevertheless, its scope is confined when applied to systems in non-equilibrium conditions, possessing long-range interactions or displaying non-extensive features. These limitations have motivated the development of generalized formalisms that extend beyond BG theory, while including the latter as a special case.

Among the most common generalizations, Tsallis statistics~\cite{Tsallis:1987eu} introduces a deformed entropic functional that accommodates non-extensive behaviors typical of large-scale systems. The resulting mechanics has attracted growing interest in the last decades, particularly due to its capability to address the area-scaling problem of black hole entropy~\cite{Tsallis:2012js}. A similar deformation occurs in Barrow model~\cite{Barrow:2020tzx}, motivated by a quantum gravitational (fractal-like) structure of the horizon geometry of black holes (see also~\cite{Saridakis:2020zol,Luciano:2022ffn} for applications to Cosmology), while a power-law corrected entropy appears in~\cite{Das:2007mj,Telali:2021jju} inspired by the idea that the origin of black-hole entropy may lie in the entanglement of quantum fields between inside and outside of the horizon. 

On the other hand, Kaniadakis ($\kappa$-deformed) statistics arises from the effort to construct a self-consistent relativistic theory~\cite{Kaniadakis:2002zz,Kaniadakis:2005zk,Kaniadakis:2009ka}. {Concretely, its foundations are tightly connected with the observation that, in special relativity, the mathematical expressions of physical observables like the momentum, the energy etc., are one-parameter (light speed) continuous deformations of the corresponding
quantities in the classical physics. In turn, Lorentz group symmetries impose a proper one-parameter deformation of the classical BG entropy too. The ensuing entropic functional allows to construct a coherent relativistic
statistical theory, which maintains the main properties (maximum entropy principle, thermodynamic stability and Lesche stability, expansivity, etc.) of the classical statistical mechanics. Deviations from the latter
are quantified by the dimensionless parameter $-1<\kappa<1$, which, in general, contains information on the relativistic features of the system. In other terms, the physical interpretation of $\kappa$ is connected with the complex (i.e. non-classical) nature of the phenomenon under investigation. 
The special limit $\kappa\rightarrow0$ recovers the classical entropy. 
In light of these considerations, it is licit to regard Kaniadakis paradigm as an effective generalization of BG entropy that naturally emerges in various high-energy contexts, such as astrophysics, cosmology, gravitation and particle physics~\cite{Kaniadakis:1996ja,Kaniadakis:1997xr}. In parallel, it is worth mentioning that alternative studies have proposed $\kappa$-theory as an approach to
epidemiology~\cite{Kaniadakis:2020ncx}, seismology~\cite{PhysRevE.89.052142}, economics~\cite{Clementi2012ANM} and natural sciences~\cite{Wada:2023hpv}, 
posing it as a serious candidate to extend the ordinary statistical mechanics into systems exhibiting non-trivial correlations~\cite{Luciano:2022eio}. }

Starting from the above premises, in the present work we 
review the impact of Kaniadakis entropy on gravitational and cosmological systems. {We shall not deal with other modified  entropies, e.g. Tsallis, Landsberg-Vedral, Sharma–Mittal, Rény and Barrow entropies (see~\cite{Tsallis:2009zex,beck2009generalised,Kaniadakis:2004rj,Barrow:2020tzx,Reny,sharma1975new,landsberg1998distributions,Luciano:2023wtx,Luciano:2021ndh} for more details and applications). However, similar considerations could be applied and expanded upon in those contexts as well.} 
On one hand, we explore how the $\kappa$-statistics affects established predictions and concepts derived from the standard BG framework. On the other hand, we shed the light on some emergent phenomena and novel behaviors that are not accounted for by the conventional theory. 
These findings indicate that Kaniadakis entropy could not only challenge the existing 
BG paradigm, but also offer profound opportunities for advancing both theoretical and applied aspects of our understanding of gravitational and cosmological environments in extreme regimes. 

The structure of the work is organized as follows: in the next Section, we review the theoretical foundations and mathematical background of Kaniadakis statistics. In Sec.~\ref{Astro}, we focus on 
astrophysical applications, while Sec.~\ref{ExGr} is dedicated to modified theories of gravity and black hole $\kappa$-thermodynamics. 
Generalized models of dark energy and effects on the history of the Universe
are discussed in Sec.~\ref{Cosmo}. 
Finally, the last section contains a summary of the work. Unless stated otherwise, we adopt Planck units $\hslash=c=k_b=G=1$.

\section{Mathematical background}
\label{MatFra}

Kaniadakis statistics naturally emerges as a one-parameter continuous deformation of the BG mechanics within Einstein special relativity~\cite{Kaniadakis:2002zz,Kaniadakis:2005zk}. It preserves the mathematical structure of the classical framework, which is in turn recovered as the deformation parameter tends to zero. The resulting theory is suitable to describe the behavior of systems exhibiting power law distributions and a large class of experimentally observed phenomena in high energy physics.

The basic ingredient of $\kappa$-statistics 
is the modified entropy~\cite{Kaniadakis:2002zz,Kaniadakis:2005zk}
\be
\label{KE}
S_{\kappa}=-\sum_{i}n_i \ln_\kappa n_i\,,
\ee
where the $\kappa$-logarithm is defined as 
\be
\label{logk}
\hspace{2.7cm}\ln_{\kappa}x\equiv\frac{x^\kappa-x^{-\kappa}}{2\kappa}\,, \qquad -1<\kappa<1\,.
\ee 
The generalized BG distribution for the $i$-th energy level $E_i$ is obtained by maximizing, after properly constrained, the
entropy~\eqref{KE}. As a result, one finds~\cite{Kaniadakis:2002zz,Kaniadakis:2005zk}
\begin{equation}
\label{ni}
n_i\equiv \exp_\kappa\left[-\beta\left(E_i-\mu\right)\right],
\end{equation}
where $\mu$ is the chemical potential, $\beta=\left(\sqrt{1-\kappa^2}\,\hspace{0.2mm}T\right)^{-1}$ the inverse temperature of the system and the $\kappa$-exponential is defined by
\begin{equation}
\ln_\kappa\left[\exp_\kappa\left(x\right)\right]=x\,\,\,\,\,\Longrightarrow\,\,\,\, \,\exp_\kappa(x)\equiv\left(\sqrt{1+\kappa^2 x^2}+\kappa x\right)^{1/\kappa}\,.
\end{equation}

\begin{figure}[t]
\centering
\captionsetup{justification=centering}
\includegraphics[width=8cm]{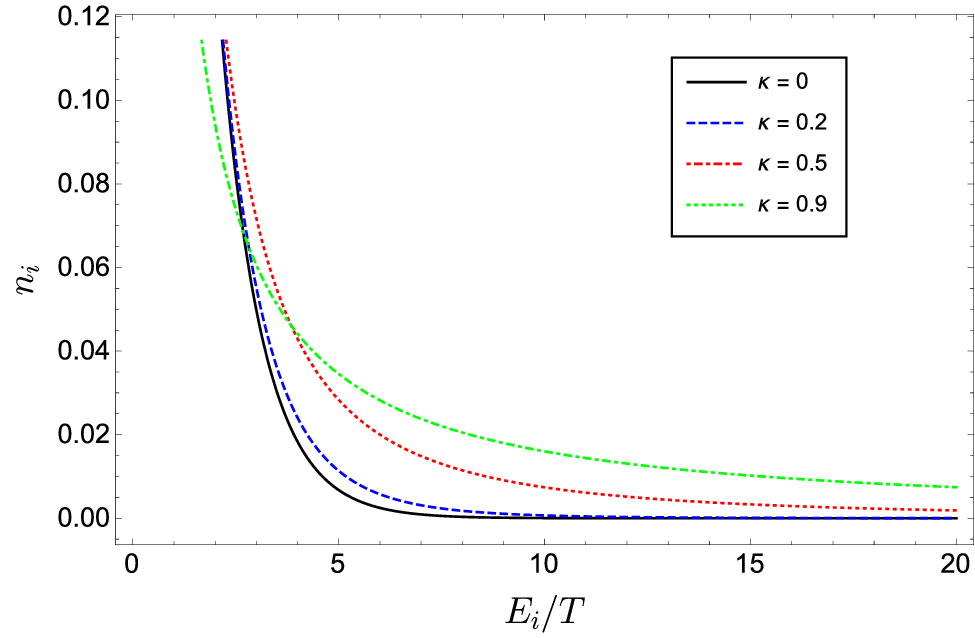}
\caption{Plot of $n_i$ in Eq.~\eqref{ni} versus $E_i/T$, for fixed $\mu=0$ and different values of $\kappa>0$.}
\label{Dist}
\end{figure}

The plot of $n_i$ versus $E_i/T$ is displayed in Fig.~\ref{Dist} for different values of $\kappa$ (since $\exp_\kappa(x)$ is even with respect to $\kappa\rightarrow-\kappa$, here and henceforth we restrict to the $\kappa>0$ domain). It can be noted that the higher the deformation parameter, the larger the departure from the Maxwell-Boltzmann distribution (black solid curve). The discrepancy becomes more evident for sufficiently large $E_i/T$, where Kaniadakis distribution has a power-law tail, as opposed to the exponential Maxwell-Boltzmann decay.

Some comments are in order here: first, it is an easy check that the laws of the classical statistics are restored in the $\kappa\rightarrow0$ limit. 
In general, $\kappa$ is not fixed by the theory itself; rather, it needs to be constrained or determined by aligning theoretical predictions  with actual experimental results. So far, the best agreement has been achieved by fitting the spectrum of cosmic rays over several decades in energy, obtaining $\kappa\sim\mathcal{O}(10^{-1})$~\cite{Kaniadakis:2002zz} (see Sec.~\ref{Astro} for details). Conversely, cosmological studies yield much tighter constraints~\cite{Luciano:2022eio}. This suggests that, although the behavior of the large-scale Universe might not deviate significantly from the classical predictions, effects of $\kappa$-entropy could still manifest through non-negligible imprints, especially in the early Universe, which is not
fully understood in the standard cosmological model.

For applications of  $\kappa$-entropy to holographic gravity~\cite{tHooft:1993dmi,Susskind:1994vu,Maldacena:1997re}, it proves convenient to recast Eq.~\eqref{KE} in the equivalent form~\cite{Abreu:2016avj,Abreu:2017fhw}
\be
\label{KanP}
S_\kappa=-\sum_{i=1}^W \frac{P_i^{1+\kappa}-P_i^{1-\kappa}}{2\kappa}\,,
\ee
where $P_i$ gives the probability of the system being in the $i$-th state and $W$ is total number of accessible states. For equiprobable configurations $P_i=1/W$, Boltzmann's entropy formula allows us to write Eq.~\eqref{KanP} in terms of the classical entropy $S=\log W$ as 
\be
\label{KenBH}
S_\kappa=\frac{1}{\kappa}\sinh \left(\kappa S\right)\,.
\ee
Since Kaniadakis corrections are constrained around $\kappa=0$, it makes sense to expand $S_\kappa$ to the leading order as 
\be
\label{Appr}
S_{\kappa}=S+\frac{\kappa^2}{6}S^3+\mathcal{O}(\kappa^4)\,.
\ee

We shall see below that this relation serves as a starting point for many applications of $\kappa$-entropy in the realm of black hole thermodynamics and cosmological scenarios.

\section{Astrophysical paradigms}
\label{Astro}

Astrophysical systems often exhibit long-range interactions and non-trivial dynamics not fitting the standard statistics. Also, they exist under extreme relativistic conditions, representing the ideal ``laboratory'' to test the correctness and predictability of $\kappa$-theory. Toward this end, we here collect recent findings on astrophysical applications of $\kappa$-statistics. The results are discussed in relation to 
challenging studies focused on constraining the deformation parameter, highlighting implications for understanding cosmic phenomena and shaping future astrophysical models.

\subsection{Cosmic rays spectrum}
As stated above, cosmic rays represent one of the most compelling systems to explore Kaniadakis paradigm. These particles are essentially the normal nuclei as in the standard cosmic abundances of
matter. Therefore, they can be thought of as an equivalent
statistical ensemble of identical relativistic particles with masses of order of the proton mass ($\sim 938\,\mathrm{MeV}$). 

It is known that the spectrum of cosmic rays, which extends over 13 decades in energy and spans 33 decades in particle flux, does not decay exponentially, violating the BG equilibrium distribution. As a matter of fact, simple calculations show that it has an approximately power-law behavior, with the spectral index depending on the energy scale. Based on this observation, an improved consistency between theoretical predictions and experimental data was found in~\cite{Kaniadakis:2002zz} by characterizing the cosmic rays spectrum through the Kaniadakis distribution
\begin{equation}
\label{distc}
    f(E)=\exp_\kappa\left(-\frac{1}{\kappa}\frac{E-\mu}{m}\right),
\end{equation}
where $m$ is the particle mass and the information on the temperature is hidden in the parameter $\kappa$. 
One can check that the asymptotic regime of $E\rightarrow\infty$ gives the approximated function $f(E)\approx\left(\frac{m}{2E}\right)^{1/\kappa}$~\cite{Kaniadakis:2002zz}, which indeed shows the correct power-law decay. 

The distribution~\eqref{distc} can be used to infer the cosmic rays flux $\Phi(E)\approx p^2f(E)$, where $p^2=E^2-m^2$ is the relativistic particle momentum. 
Following~\cite{Kaniadakis:2002zz}, computations are easily done by taking into account the relativistic expression
relating $E$ and $p$, yielding
\be
\label{fluxc}
\Phi(E)= A\left[\left(\frac{E}{m}+1\right)^2-1\right]\exp_{\kappa}\left[-\beta\left(E-\mu\right)\right]\,,
\ee
with $\beta$ being related to the inverse temperature. It is a direct check that this expression, in agreement with observations, decays following the power-law behavior $\Phi(E)\propto E^{-a}$, where $a=\sqrt{1+\left(\frac{m}{T}\right)^2}-2$.
In particular, 
comparison with BG statistics reveals that Kaniadakis framework is better suited to fit the observed cosmic rays flux for $A=10^5 \,(\mathrm{m^2\,\hspace{0.4mm} sr\,\hspace{0.4mm} s\,\hspace{0.4mm} GeV})^{-1}$, $mc^2=938\,\mathrm{MeV}$, $\mu=-375\,\mathrm{MeV}$ and $k_B T=208\,\mathrm{MeV}$, provided that $\kappa=0.2165$~\cite{Kaniadakis:2002zz}.\footnote{We have here restored full units for direct comparison with~\cite{Kaniadakis:2002zz}.}

At this point, it is interesting to remark that the power law asymptotic behavior of Eq.~\eqref{fluxc} is due to the presence of the $\kappa$-exponential, which, in turn, can be naturally traced to the composition law of momenta in special relativity through a suitable redefinition of the sum in the underlying algebraic structure ($\kappa$ sum~\cite{Kaniadakis:2002zz}). In other terms, such a power law behavior of the cosmic flux can be regarded as a signature of the particle relativistic nature.
Such a result provides strong evidence that the intricate nature of relativistic systems demands a reevaluation of the fundamentals of the classical statistical mechanics. From this perspective, 
Kaniadakis model stands out as a promising candidate to improve the current limits of BG theory in the relativistic regime.

\subsection{Open Stellar clusters}
\label{StCl}

Stellar clusters are groups of stars held together by gravity. There are two main types of such systems: globular clusters, which are large and spherical, containing thousands to millions of stars, and open clusters, which are less dense, with roughly hundreds to thousands of stars. Open clusters, often located in the disk of galaxies like the Milky Way, are younger and less tightly bound than globular clusters. 

Because of their large number of constituents, open clusters obey statistical laws that govern their collective properties and dynamics. There are three dominant mechanisms
underlying the long-term evolution of these systems:

\begin{itemize}
\item[$\bullet$] \emph{collisional relaxation}, which occurs when stars within a cluster undergo close encounters or collisions, transferring energy and momentum among themselves. Over time, such interactions lead to a uniform distribution of velocities among stars, ultimately driving the cluster towards a more dynamically relaxed state (see, for instance,~\cite{Frenk:1995fa} and references therein).

\vspace{3mm}

\item[$\bullet$]\emph{Lynden-Bell relaxation}, which refers to the exchange of energy due to gravitational interactions. This process plays a central role in adjusting the velocities of stars within the cluster, leading to a more isotropic distribution~\cite{Lynden-Bell:1966zjv}.  

\vspace{3mm}

\item[$\bullet$] \emph{Relaxation associated with radial orbit instability}, which pertains specifically to the instability of stars due their gravitational interactions. It involves the expansion of orbits and, in some cases, the escape of the most energetic stars.
This process contributes to the gradual dissolution of open clusters over cosmic timescales~\cite{Frid}. 
\end{itemize}

If, on one hand, the first two mechanisms are adequately described by the standard statistical mechanics, 
orbital relaxation has not yet been fully understood. In this sense, a tentative solution was provided in~\cite{2010EL9169002C}, introducing the Kaniadakis distribution 
\begin{equation}
\label{resvel}
\phi_\kappa(v_r)\propto\exp_{\kappa}\left(-\frac{v_r^2}{\sigma^2}\right)
\end{equation}
for the stellar residual radial velocity $v_r$ 
in some baseline open clusters. Application of the Kolmogorov-Smirnov statistical test~\cite{ref1} for each observed cumulative distribution showed that Eq.~\eqref{resvel} aligns with observations more accurately than the Maxwell-Boltzmann exponential, assuming a non-trivial dependence of $\kappa$ on stellar-cluster ages. In particular, for some selected stellar clusters (e.g., Pleiades, NGC 1039, NGC 6475, NGC 2099, NGC 6633, Praesepe etc.), the entropic parameter is estimated around unity, while a slightly lower value $\kappa\sim\mathcal{O}(10^{-1})$ is obtained for the NGC 0752 system~\cite{2010EL9169002C}.

\subsection{Jeans instability criterion}
The Jeans instability serves as a criterion to ascertain the conditions for gravitational instability and subsequent collapse in clouds of interstellar gas. In its basic form, it states that a region of thermal gas will collapse under its self-gravity if its size exceeds the critical length  
\be
\label{lambda}
\lambda_j=\sqrt{\frac{\pi T}{\mu m_H \rho_0}}\,,
\end{equation}
where $T$ is the temperature of the gas, $\mu$ the mean molecular weight, $m_H$ the mass of a hydrogen atom and $\rho_0$ the equilibrium mass density of the system, respectively~\cite{alma991002292379703816}. 

To conceptualize the Jeans length, the easiest way is to rewrite approximately $\rho_0$ as the mass-to-volume ratio of the cloud, i.e. $\rho_0=M/r^3$. Dropping irrelevant numerical factors, Eq.~\eqref{lambda} becomes $\lambda_j\propto\sqrt{\frac{T r^3}{\mu M}}$. Hence, when $r=\lambda_j$, we have
$T=M\mu/r$, i.e. the thermal energy stemming 
from the agitation of particles
equals the self-gravitational energy and the cloud is in equilibrium. 
Conversely, all scales larger than $\lambda_j$ are unstable to gravitational collapse, whereas smaller scales are stable.

The critical length~\eqref{lambda} is consistent with the classical equipartition theorem. Nevertheless, due to the relativistic nature of interstellar gas constituents, a derivation within Kaniadakis kinetic theory seems all the more natural. This was investigated in~\cite{Abreu:2016avj}, obtaining the modified expression
\begin{equation}
\label{Jeansmod}
\lambda_c^\kappa=\sqrt{\frac{\left(1+\frac{\kappa}{2}\right)}{\left(1+\frac{3}{2}\kappa\right)2\kappa}\frac{\Gamma\left(\frac{1}{2\kappa}-\frac{3}{4}\right)\Gamma\left(\frac{1}{2\kappa}+\frac{1}{4}\right)}{\Gamma\left(\frac{1}{2\kappa}+\frac{3}{4}\right)\Gamma\left(\frac{1}{2\kappa}-\frac{1}{4}\right)}}\,\lambda_J\,,
\end{equation}
where $\Gamma(x)$ denotes the Euler Gamma function. It is important to observe that, for $\kappa=0$, this expression gives the correct limit $\lambda_c^{\kappa}\rightarrow\lambda_J$, while considering $0\le\kappa<2/3$ results in $\lambda_c^\kappa> \lambda_j$. Intuitively, relativistic statistical corrections
contribute to increasing the thermal kinetic energy of the cloud, thus making the system more stable against gravitational perturbations.  On the other hand, in the limit case $\kappa=2/3$, the $\kappa$-generalized equipartition theorem suffers from a divergence and the reasoning below the above derivation may need to be reconsidered~\cite{Abreu:2016avj}.

Further studies of Jeans instability appeared within the framework of gravity beyond general relativity (see Sec.~\ref{ExGr} for more discussion). Specifically, in~\cite{Yang:2020ria} a dispersion relation generalizing the Jeans modes was derived in Eddington-inspired Born-Infeld model by deforming the Maxwell-Boltzmann function to a family of power law distributions. Likewise, exploiting the kinetic theory, the analysis of~\cite{He_2022} explored the influence of Kaniadakis statistics in the background of $f(R)$ gravity, which is a family of theories that extend the Lagrangian of the Einstein–Hilbert action by including suitable functions of the Ricci scalar $R$.
The main outcome was that the range of the unstable modes and the growth rates of density perturbations decrease as the parameter $\kappa$ increases in both the high and low frequency regime. 

The behaviors of Jeans modes for gravitational systems composed of dark and baryonic matters were the subject of investigation in~\cite{YXC}. The results showed that the Jeans instability is suppressed due to 
relativistic corrections, setting the conditions for new phenomenological models in cosmology.  

\section{Extended Gravity and Black hole thermodynamics}
\label{ExGr}

Einstein's theory of General Relativity is the cornerstone of gravitational physics. Its success in explaining a variety of phenomena stands as a testament of its robustness. However, challenges arise when attempting to unify this framework with quantum mechanics.  Such discrepancies highlight the need for a more comprehensive formalism, capable of describing gravity at both macroscopic and microscopic scales. 

In what follows we discuss the interplay between Kaniadakis theory and extended approaches to gravity. These studies involve exploring to what extent non-extensive statistical principles could shed light on 
unexplained phenomena, especially in scenarios involving the extreme gravitational conditions of black holes environments and the very early Universe.  
 
\subsection{Verlinde's entropic gravity}
\label{Verlinde}

Verlinde's model is among the most engaging approaches that seek to connect the concepts of entropy and gravity~\cite{Verlinde:2010hp}. In a nutshell, it suggests that gravity might not be a fundamental force, as described in Einstein's theory, but rather an outcome of the way microscopic information is encoded on the surface of bodies. From this perspective, the gravitational force could emerge from changes in the entropy associated with these information-bearing boundaries.

Verlinde's idea was used in~\cite{Abreu:2017fhw} along with Kaniadakis statistical principles to derive an effective Newton's constant in the form
\begin{equation}
\label{Gmod}
G_{\kappa}=\frac{\left(1+\frac{3}{2}\kappa\right)2\kappa}{\left(1+\frac{\kappa}{2}\right)}\, \frac{\Gamma\left(\frac{1}{2\kappa}+\frac{3}{4}\right)\Gamma\left(\frac{1}{2\kappa}-\frac{1}{4}\right)}{\Gamma\left(\frac{1}{2\kappa}-\frac{3}{4}\right)\Gamma\left(\frac{1}{2\kappa}+\frac{1}{4}\right)}\,G\,,
\end{equation}
for $0\le\kappa<2/3$, where we have restored the gravitational constant $G$ for the sake of clarity. 

\begin{figure}[t]
\centering
\captionsetup{justification=centering}
\includegraphics[width=8.5cm]{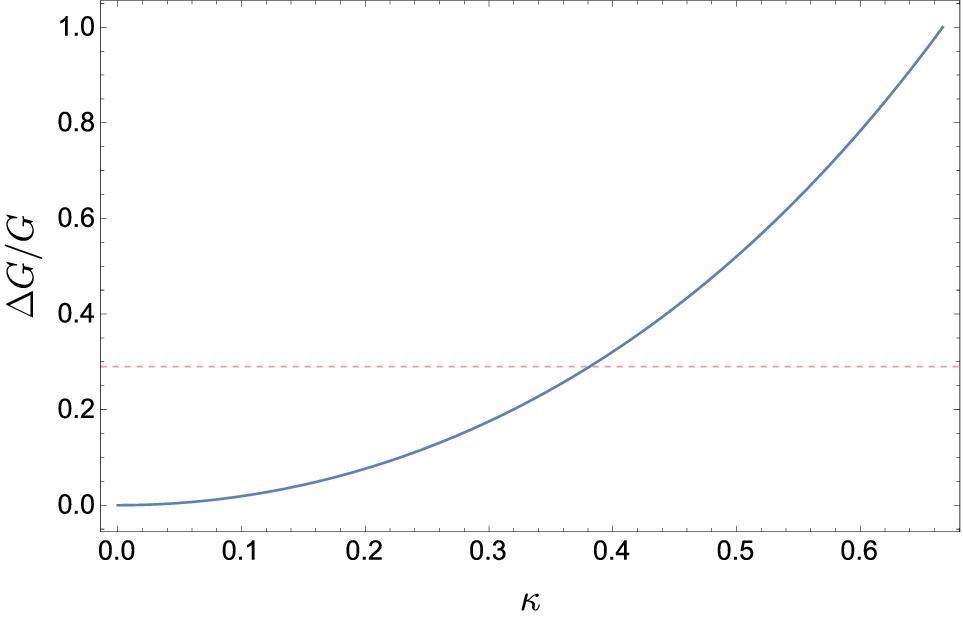}
\caption{Plot of $\Delta G/G$ versus $\kappa$. The red dashed line  marks the observational bound $|\Delta G/G|\lesssim0.29$~\cite{Umezu:2005ee,Nesseris:2006jc,Verbiest:2008gy,Nesseris:2017vor,Asimakis:2022mbe}.}
\label{Gk}
\end{figure}

In Fig.~\ref{Gk} we plot the relative difference $\frac{\Delta G}{G}$ versus $\kappa$, where $\Delta G\equiv G-G_\kappa$. 
It can be seen that $G_\kappa\le G$ for any allowed $\kappa$, with the equality being satisfied as $\kappa\rightarrow0$. 
Such a behavior entails that, for a given source, the intensity of  Verlinde's gravity in Kaniadakis framework is reduced comparing to the classical theory. Notice that this is an alternative way to convey the result in Eq.~\eqref{Jeansmod}.

The red dashed line in Fig.~\ref{Gk} marks the observational constraint $|\frac{\Delta G}{G}|\lesssim0.29$, which is obtained by cosmological constraints on large-scales~\cite{Umezu:2005ee}, and SNIa~\cite{Nesseris:2006jc}, PSR J0437-4715~\cite{Verbiest:2008gy} and Planck~\cite{Nesseris:2017vor} data in modified gravity frameworks. Hence, the constraint $\kappa\lesssim0.38$ is obtained on Kaniadakis parameter. Notice that this is of the same order as the constraint from cosmic rays spectrum, although it is much less stringent than other bounds of cosmological origin (see Sec.~\ref{Cosmo}).

Also, it is interesting to observe that a smaller gravitational constant (of course within the observational bounds) 
is known to be one of the mechanisms
that can alleviate the $\sigma_8$ tension, where $\sigma_8$ is the standard deviation of the density fluctuation in an 8 $h^{-1}\,\mathrm{Mpc}$ radius sphere (see Sec.~\ref{Cosmo} for more discussion). Thus, Kaniadakis Cosmology may provide an effective framework to explain this tension.

A related outcome is that self-gravitating systems will take longer to collapse. This follows directly from the computation of the free fall time, which in $\kappa$-statistics takes the generalized form~\cite{Abreu:2017fhw}
\begin{equation}
\label{tff}
t_{FF}^\kappa=\sqrt{\frac{\left(1+\frac{\kappa}{2}\right)}{\left(1+\frac{3}{2}\kappa\right)2\kappa}\,\frac{\Gamma\left(\frac{1}{2\kappa}-\frac{3}{4}\right)\Gamma\left(\frac{1}{2\kappa}+\frac{1}{4}\right)}{\Gamma\left(\frac{1}{2\kappa}+\frac{3}{4}\right)\Gamma\left(\frac{1}{2\kappa}-\frac{1}{4}\right)}}\, t_{FF}\,,
\end{equation}
where $t_{FF}=\sqrt{\frac{3}{2\pi\rho_0}}$ is the standard expression. Since the $\kappa$-dependent factor is 
the reciprocal square root of the correction in Eq.~\eqref{Gmod}, 
one has $t^\kappa_{FF}\ge t_{FF}$, where the inequality is obviously saturated for $\kappa\rightarrow0$.

\subsection{Loop quantum gravity}
\label{LQG]}

Loop Quantum Gravity (LQG) provides a theoretical attempt to unify quantum mechanics and Einstein's theory of gravity. A distinctive feature of this approach is that geometrical quantities like area and volume are described by quantum operators with discrete spectra. This reflects the expected granular structure in the fabric of quantum spacetime, which is common to many models of quantum gravity~\cite{Addazi:2021xuf,Bosso:2023aht}. Concretely, the size of the quantum of area in Planck units is measured by the Immirzi parameter $\gamma=\log 2/\left(\pi\sqrt{3}\right)$, which is estimated by counting the number of microstates of a given system classically~\cite{Immirzi:1996dr} (see also~\cite{Ashtekar:2000eq} for more physical insights on the meaning of $\gamma$). 

In~\cite{Abreu:2018mti} Abreu \emph{et al.} worked on a generalization of the Immirzi parameter to non-extensive statistics. For a system of boundary surface $A$ in the microcanonical ensemble, they obtained a new relation between the Immirzi parameter, the $\kappa$ parameter and the surface area in the form
\begin{equation}
\gamma_\kappa=\gamma\, \frac{A}{4\log\left[\exp_{\kappa}\left(A/4\right)\right]}\,,
\label{Imm}    
\end{equation}
which implies $\gamma_\kappa>\gamma$, and $\gamma_\kappa\rightarrow\gamma$ for vanishing $\kappa$, as expected. Therefore,  Kaniadakis corrections contribute to modify, and in particular to increase the size of the quantum of area, revealing a potential role of $\kappa$-statistics in the quest for a quantum gravity formalism. 

\subsection{Modified gravity and expanding Universe}
\label{DEGr}

The success of the $\Lambda$-CDM model arises from its remarkable match with the available cosmological observations. However, a number of issues challenge its ability to describe the evolution of the Universe at both infrared and ultraviolet scale. In this gap, modified cosmology based on extended gravity provides one of the most fruitful paradigms to improve our current knowledge of the Universe. 

Based on a generalized model of dark energy with $\kappa$-entropy, 
the analysis in~\cite{Jawad:2023aog} explored the cosmic evolution within the frameworks of $f(G)$ and $\tilde f(T)$ gravity, where $G$ and $T$ are the Gauss–Bonnet invariant and torsion scalar, respectively. Using these models, the behavior of various cosmic parameters was examined to discuss the accelerated expansion of the Universe. Similar investigation was conducted in~\cite{Ghaffari:2021xja,Sania:2023fjx} in Brans–Dicke gravity, which extends Einstein's theory by introduction of a scalar field interacting with gravity.
Such studies fit into a very active research trend that aims to scrutinize the statistical properties behind the concept of dark energy (see Sec.~\ref{Cosmo} for more discussion).

\subsection{Black hole thermodynamics}
\label{BH}

Thermodynamics of black holes provides a unique arena for testing the interplay between gravity, relativity and quantum theory. This domain emerges by conceptualizing black holes as thermodynamic systems, revealing intriguing parallels between their properties and classical thermodynamics~\cite{Bekenstein:1973ur,Bekenstein:1974ax,Hawking:1974rv,Hawking:1975vcx}. In this framework, an important concept is black hole entropy. Arguments from different perspectives, including the holographic principle\footnote{The holographic principle is a key concept in quantum gravity. It posits that  the information within a defined $d$-dim. space can be fully encoded on a $(d-1)$-dim. surface surrounding that space.}~\cite{tHooft:1993dmi,Susskind:1994vu,Maldacena:1997re}, suggest that the entropy of a black hole should obey the Bekenstein-Hawking area law $S_{BH}=A/4$, where $A$ is the horizon surface area~\cite{Bekenstein:1973ur}. However, it is clear that, if we consider black holes as genuinely $3$-dim. systems, $S_{BH}$ cannot be identified as thermodynamic entropy, since it violates extensivity~\cite{Tsallis:2012js}. 

\begin{figure}[t]
\centering
\includegraphics[width=8cm]{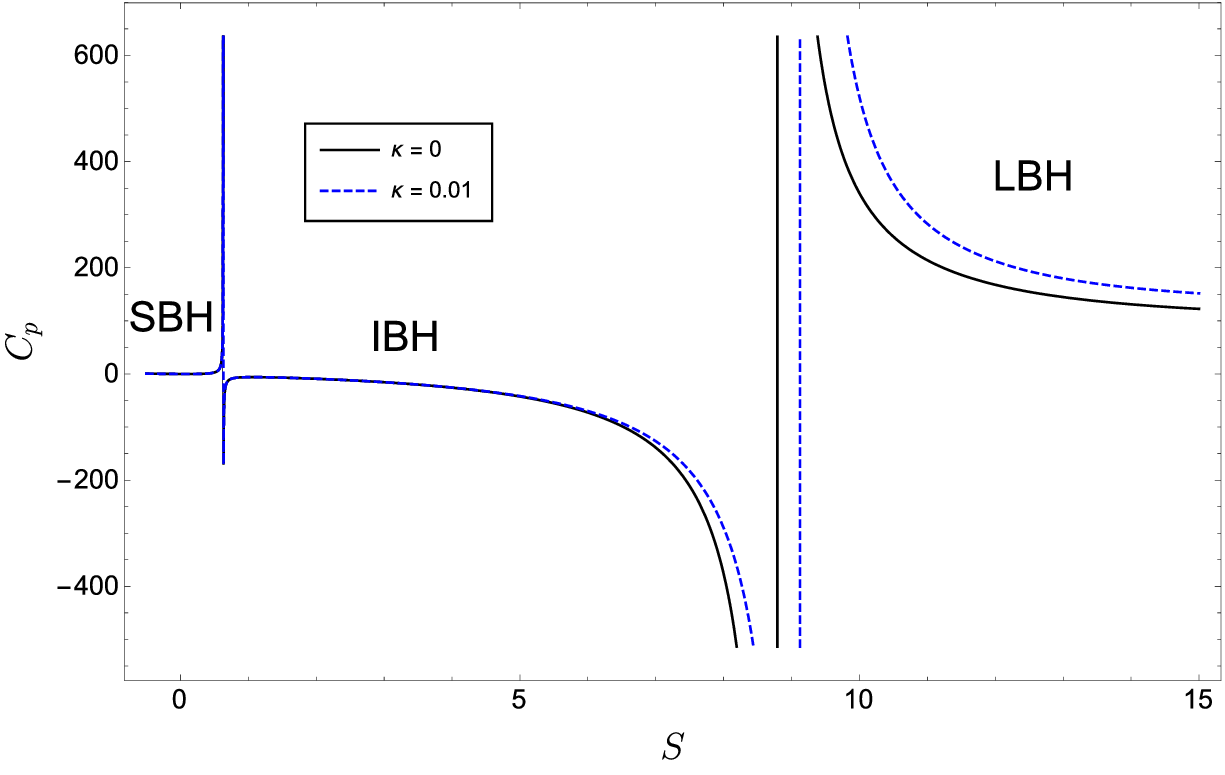}
\caption{Plot of the heat capacity at constant pressure versus entropy. SBH and LBH denote the ``small'' and ``large'' stable (i.e. $C_p>0$) domains, respectively, while IBH is the ``intermediate'' unstable (i.e. $C_p<0$) region (from~\cite{Luciano23}).}
\label{heatca}
\end{figure}

To understand this puzzle, the thermodynamics of black holes was generalized to accommodate non-extensive definitions of entropy~\cite{Luciano:2023fyr,Luciano23,Odintsov:2023qfj,Cimidiker:2023kle}.  Interesting features were found in the context of thermodynamics based on Eq.~\eqref{KenBH}, showing that relativistic $\kappa$-entropy has a non-trivial impact onto the phase structure and transitions of charged anti-de Sitter black holes~\cite{Luciano:2023fyr}. For concreteness, we display in Fig.~\ref{heatca} the behavior of the heat capacity at constant pressure versus the entropy. Since divergences of $C_p$ are associated to (first-order) small-to-large phase transitions of black holes, it is evident that Kaniadakis corrections affect the black hole size at which such critical phenomena occur. Further study of the $P-V$ diagram and Gibbs free energy highlighted a phenomenology similar to that of van der Waals fluid~\cite{Luciano:2023fyr}, with the critical exponents being now dependent on $\kappa$.

Along the same line, Cimidiker \emph{et al.} 
analyzed the effects of a quantum gravitational deformation of the Heisenberg Principle on non-extensive black hole thermodynamics~\cite{Cimidiker:2023kle}. As a result, it was shown that the final state of a black hole at the end of the evaporation turns out to be a remnant with finite temperature and entropy. Apart from its own interest, this achievement could be exploited to tackle the black hole information paradox in the future~\cite{Inprepa}.

\section{Cosmological insights}
\label{Cosmo}

The observation that the Universe is currently accelerating presents a significant conundrum in cosmology. To account for this phenomenon, the most plausible mechanism is the existence of an unknown form of energy - the \emph{dark energy} (DE) - which would permeate all of space. Because of its negative pressure, 
DE would exert a repulsive gravitational force, providing the fuel to drive the accelerated cosmic expansion. Although DE is thought to make up about 70\% of the energy budget of the Universe, its properties remain poorly understood. 

Since its theorization in the late 1990s, DE 
has been extensively studied~\cite{Copeland:2006wr}. Among the various models proposed so far, holographic dark energy (HDE) has a central role in quantum gravity, as it is founded on the application of the holographic principle to the horizon degrees of the freedom of the Universe~\cite{Wang:2016och}. In this model,  the energy density of DE takes the form
\begin{equation}
\label{DE}
   \rho_{de}=3C^2M_p^2L^{-2}\,,
\end{equation}
where $C$ is a constant of order unity and $L$ a characteristic length scale that acts as IR cutoff (we have here restored the Planck mass $M_p$ for clarity).  The choice of the future event horizon $L=R_h=a\int_t^\infty \frac{dt}{a}$ was proven to be in line with phenomenology~\cite{Wang:2016och}, though improvements and further analysis could be needed~\cite{Li:2008zq}.

\subsection{Kaniadakis modification of Friedmann cosmology}
\label{KHDE}

Let us elaborate on implications of $\kappa$-entropy for cosmology. We focus on the generalization of Eq.~\eqref{DE} and the predictions that emerge from the modified Friedmann equations governing the Universe's evolution.

\subsubsection{Kaniadakis holographic dark energy}

The relation~\eqref{DE} is 
framed in the semiclassical theory, where the Universe's entropy is described by the usual 
Bekenstein-Hawking area law $S_{BH}$ (see Sec.~\ref{BH}). On the other hand, a generalized HDE model was proposed in~\cite{Drepanou:2021jiv}, using the entropy~\eqref{Appr} and the future horizon as IR cutoff\footnote{Other descriptions of KHDE set the apparent horizon as IR cutoff~\cite{Moradpour:2020dfm,Jawad:2021xsr,Sharma:2021zjx}. Besides not getting back  the standard HDE and thermodynamics as a sub-case, the disadvantage of these models is that they need unacceptably large values of $\kappa$ to fit observational results.}. In the ensuing framework, the energy density of Kaniadakis Holographic Dark Energy (KHDE) reads
\begin{equation}
\label{KHDEde}
    \rho_{de}^\kappa=3C^2M_p^2R_h^{-2}+\kappa^2 M_p^6R_h^2\,.
\end{equation}

Based on this relation, modified Friedmann equations were derived that reproduce the observed history of the Universe~\cite{Drepanou:2021jiv} (see also~\cite{Sheykhi:2023aqa} for an alternative derivation). This model also allowed to compute characteristic quantities like the equation of state parameter of KHDE, the deceleration parameter, the squared sound speed and the Hubble parameter. 
Experimental implications of KHDE were further addressed in~\cite{Hernandez-Almada:2021aiw,Hernandez-Almada:2021rjs,Singh:2022ubm,Luciano:2022knb}. Notably, the analysis of~\cite{Hernandez-Almada:2021rjs} revealed that Kaniadakis horizon-entropy model
may alleviate the Hubble tension problem, representing a potentially successful candidate for the description of the cosmic evolution. Further study was conducted in~\cite{Sheykhi:2023aqa} by investigating the validity of the generalized second law of thermodynamics for the Universe enclosed by the apparent horizon. The obtained equations provide a background to explore a new cosmological model built on the $\kappa$-entropy, which might be used to study the evolution of the Universe from early deceleration to the late-time acceleration and constrain the $\kappa$-parameter accordingly. In this context, the analysis of~\cite{Hernandez-Almada:2021rjs} reveals that $|\beta|\equiv|\frac{\kappa\pi}{G H_0^2}|\lesssim\mathcal{O}(10^{-3}\div10^{-2})$ for consistency with the current observed value of the deceleration parameter $q_0$ in the presence of a non-vanishing cosmological constant $\Lambda$. In turn, this entails the very stringent constraint $|\kappa|\lesssim\mathcal{O}(10^{-126}\div10^{-125})$, where we have set the present Hubble rate $H_0\simeq10^{-42}\,\mathrm{GeV}$. Also, the dynamical system analysis of~\cite{Hernandez-Almada:2021rjs} shows that, in a Kaniadakis-entropy based evolutionary scenario, the Universe 
past attractor is the matter-dominated epoch, while at late times, the
Universe is dominated by dark energy and undergoes an accelerated expansion in both vanishing and non-vanishing $\Lambda$ cases. The requested deviation from the standard Bekenstein-Hawking-like horizon scaling is still quantified by $|\kappa|\simeq 10^{-125}$ for $\Lambda\neq0$ and $|\kappa|\simeq 10^{-123}$ for $\Lambda=0$. The latter less stringent constraint can be understood by observing that, in the absence of an explicit cosmological constant term, one needs a more significant deviation from standard
Cosmology to describe the late-time Universe acceleration. 

While being more stringent than the astrophysical bounds discussed in Sec.~\ref{Astro}, the above cosmological constraints align with 
those inferred from other set of cosmic measurements, e.g. Observational Hubble data, Pantheon supernova Type Ia sample, H II galaxies, Strong lensing systems and Baryon acoustic oscillations (see~\cite{Hernandez-Almada:2021rjs} for more details). Therefore, it is expected that allowing even small deviations from the classical horizon entropy might lead to a very interesting cosmological phenomenology, which is in good agreement
with observational behaviour and can solve some of the current tensions of the $\Lambda$CDM model.

\subsubsection{Baryogenesis}

In~\cite{Luciano:2022knb} $\kappa$-modified Friedmann equations provided a possible solution to baryogenesis, which is the observed imbalance between matter and antimatter in the Universe. In order for this asymmetry to occur, the following three Sakharov's conditions must be satisfied: \emph{i}) violation of the baryon number; \emph{ii}) loss of thermal equilibrium; \emph{iii}) violation of C and CP symmetries~\cite{Riotto:1999yt}. While the first and last requirements can be fulfilled by introducing a suitable coupling between the spacetime and baryon current, there is no mechanism to break the thermal equilibrium in the radiation-dominated era in the standard cosmology. Interestingly enough, in~\cite{Luciano:2022knb} Kaniadakis effects were found to generate fluctuations in the energy and pressure content of the Universe, providing the right conditions to satisfy all three Sakharov's criteria. Observational consistency set $\kappa\simeq10^{-10}$. Despite being much smaller than the constraint from the cosmic rays spectrum, the non-vanishing value of $\kappa$ signals the need of a Kaniadakis-like modification of Friedmann cosmology to trigger baryogenesis.

In~\cite{Salehi:2023zqg} a relation was established between
the quantities related with the energy content
of the Universe and the cosmographic parameters that express its geometry. Reconstructing the evolution of the deceleration parameter versus the redshift, 
the model showed that Kaniadakis corrections could drive the current cosmic acceleration without introducing any ad hoc component of DE or cosmological constant.

\subsubsection{IceCube PeV neutrinos}
The IceCube Neutrino experiment was designed to detect neutrino sources in the TeV regime~\cite{IceCube:2016zyt}. Unexpectedly, the discovery of high-energy astrophysical neutrino flux with energies of the order of PeV
opened new scenarios in astroparticle physics, 
due to the difficulties to reconcile the candidate sources of these particles
with the standard cosmological model. This tension was alleviated in~\cite{Blasone:2023yke}, assuming that the Universe's evolution 
obeys $\kappa$-modified Friedmann equations with the parameter $\kappa\simeq10^{-37}$. The discrepancy with   
constraints of different origin served as inspiration for an effective dynamical model, where relativistic statistical corrections might become relevant in the very early Universe, while they gradually turn off approaching the present time.

\subsection{Inflation and perturbation growth}
Inflationary cosmology proposes that the Universe underwent a rapid and exponential expansion immediately after the Big Bang~\cite{Guth:1980zm,Starobinsky:1980te,Linde:1981mu}. This model was introduced to solve several key problems (e.g., flatness, horizon, structure formation) in standard cosmology and explain the uniformity of the cosmic microwave background radiation. Although inflation is largely accepted, the underlying mechanism and its complete implications remain areas of ongoing research and debate.

\begin{figure}[t]
\centering
\captionsetup{justification=centering}
\includegraphics[width=8.6cm]{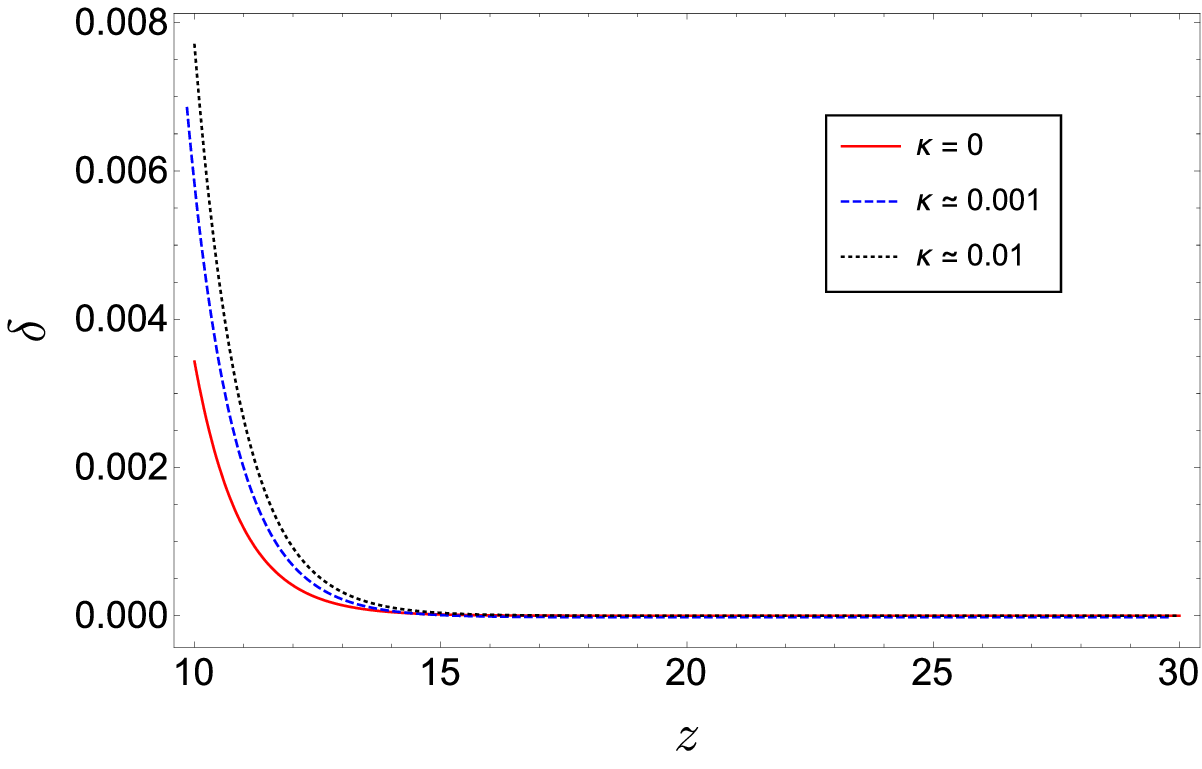}
\caption{Plot of $\delta$ versus the redshift $z$ (from~\cite{Lambiase:2023ryq}).}
\label{Fig3}
\end{figure}

In the framework of $\kappa$-modified cosmology, inflation driven by a scalar field was analyzed in~\cite{Lambiase:2023ryq} from the slow-roll condition perspective. The model was successfully tested by comparing the predicted scalar spectral index and tensor-to-scalar ratio to the latest BICEP and Planck data. Furthermore, it was proven that Kaniadakis corrections affect the growth of 
primordial fluctuations, which are the seeds of all structures in the Universe. To show this, the relevant quantity to consider
is the density contrast $\delta\equiv\delta\rho/\rho$, where $\rho$ and $\delta \rho$ are the energy density and fluctuations of the fluid that permeates the Universe, respectively. Assuming $\delta\ll1$ and the approximation~\eqref{Appr}, the following modified evolution equation was obtained 
\be
\label{dif}
\delta''+\frac{3}{2a}\left(1+\frac{9\kappa^2}{64 \rho^2}\right)\delta'-\frac{3}{2a^2}\left(1+\frac{9\kappa^2}{32 \rho^2}\right)\delta=0\,,
\ee
where the prime denotes derivative with respect to the scale factor $a$. 
The above relation was solved numerically. The solution is plotted in Fig.~\ref{Fig3}, which shows that higher values of $\kappa$ correspond to a faster perturbation growth comparing to the standard case (red solid curve). Therefore,  even very tiny deviations from classical statistics at microscopic level may result in 
appreciable effects at very large scales in the Universe.

Recently, the Universe's evolution from inflation to reheating was studied in~\cite{Odintsov:2023vpj}, using a more generalized entropy that includes Kaniadakis functional as a limit case. In that context, the inflaton field driving the accelerated expansion was represented by the entropic energy density. For suitable ranges of the entropic parameters, the Universe was found to enter to a reheating stage after the end of inflation, when the entropic energy decays to relativistic particles with a certain decay rate. On top of that, consistency was achieved between 
predicted values of the observable inflation indices and the recent Planck data.

It is worth noting that the study of the evolution
equation of matter overdensity $\delta$ is intimately connected with the $\sigma_8$ tension, which is indeed related to the matter clustering and the fact that the Cosmic Microwave Background
(CMB) estimation~\cite{Planck:2018vyg} differs from the SDSS/BOSS direct measurements~\cite{eBOSS:2018cab,BOSS:2016wmc}. In~\cite{Heisenberg:2022gqk,Heisenberg:2022lob} it was argued that an increased friction term in the matter-perturbation evolution (the second one in Eq.~\eqref{dif}) may lead to smaller $\sigma_8$, providing a potential effective mechanism to alleviate the $\sigma_8$ tension. 
Clearly, a more detailed analysis in the present framework requires investigating the behavior of 
$f\sigma_8(z)=f(z)\sigma(z)\equiv -\frac{d\log\delta(z)}{d\log z}\,\sigma_8\frac{\delta(z)}{\delta(0)}$  against the $\Lambda$CDM scenario. More work in this direction is reserved for the future. In passing, we mention that the possibility of alleviating both the $H_0$ and $\sigma_8$ tension in modified entropic scenarios was recently explored in~\cite{Basilakos:2023kvk} in the context of Tsallis Cosmology.

\subsection{Cosmological phase transitions}

Inspired by similarities with the critical phenomena of anti-de Sitter black holes, the efforts in~\cite{Housset:2023jcm} were focused on the thermodynamic phase structure of a Friedmann–Lema\^itre–Robertson–Walker Universe in Kaniadakis statistics. In particular, the study of the equation of state pointed out the existence of non-trivial critical points, where a van der Waals-like phase transition occurs. Although results were obtained 
in the leading order approximation~\eqref{Appr}, they provide the basis for a reliable thermodynamic picture of the early Universe, at least in the stable regions.

\section{Discussion and Conclusions}
Kaniadakis theory is a one-parameter extension of the traditional statistics that incorporates relativistic effects. It is based on the idea that
in extreme conditions, such as high energies or densities, standard statistical models might not accurately describe the behavior of systems, requiring suitable modifications. The effort of  
this work was to review recent advances of Kaniadakis theory in extreme gravitational and cosmological environments. The advantage of applying $\kappa$-framework to such systems is that they may somehow amplify modified entropy effects, creating suitable conditions for their direct or indirect detection. 

In the first part of the work, we discussed 
astrophysical implications on open stellar clusters, Jeans instability and gravitational collapse. We showed that Kaniadakis statistics has a non-trivial impact on the classical predictions of Jeans length and collapsing time, as it contributes to making gravitational systems more stable. Afterwards, the focus was shifted onto the interplay between Kaniadakis theory and
modified gravity, with special attention to Verlinde's formalism and Loop Quantum Gravity. 
We also explored black hole $\kappa$-thermodynamics, which exhibits a richer phenomenology compared to the ordinary description. As a final study, we investigated Kaniadakis modification of Friedmann cosmological equations. Remarkably, this framework allows for alleviating tensions and explaining phenomena that cannot be understood in the  classical theory. Hence, it stands out as a potential candidate to improve over the current limits of the standard cosmological model. Clearly, the class of phenomena and predictions to explore is too vast to cover in this review. More applications and insights of $\kappa$-theory are discussed in~\cite{Luciano:2022eio}. 
 
Despite many virtues listed above, some aspects are yet to be clarified or fully understood. As emphasized in Sec.~\ref{MatFra}, Kaniadakis theory is framed in the context of special relativity. The question naturally arises of how it can be adapted to accommodate the symmetries of general relativity. The issue of generality of $\kappa$-statistics is also tackled in~\cite{Alves:2023ttj} from the condensed matter perspective. {Regrettably, Kaniadakis paradigm fails to provide a non-extensive version of Ising model, unlike Tsallis’s distribution, which instead stands out as a potentially valid generalization (see~\cite{Carvalho:2023jqj,Alves:2023ttj,Carvalho:2022hca} for further discussion on the subject).} 
This drawback apparently challenges the role of $\kappa$-theory as generalized statistical mechanics, at least in low-energy regimes. Toward developing a more comprehensive framework, valuable insights could be gained by working with a multiparameter functional that
gathers all the deformed entropies introduced so far under a single umbrella. 
Prospective analysis appears in~\cite{Odintsov:2022qnn,Nojiri:2023bom,Odintsov:2023qfj} in cosmology and black holes thermodynamics, but further applications are under active consideration in gravity and information theory. For instance, a challenging task is to look for footprints in the spectrum of primordial gravitational waves, which carry unique knowledge on the Universe's initial conditions. On the other hand, based on~\cite{Ourabah_2015}, the use of non-extensive entropy as a candidate for a generalized quantum information theory could be explored for the purpose of solving some paradoxes in quantum gravity~\cite{Almheiri:2020cfm}. 
Work is in progress along these directions and will be presented elsewhere~\cite{Inprepa}.

\bmhead{Acknowledgments}
The author would like to acknowledge the anonymous Referees for their comments and recommendations, which contributed to improve the quality of the manuscript.
He is also grateful to the Spanish ``Ministerio de Universidades''
for the awarded Maria Zambrano fellowship and funding received
from the European Union - NextGenerationEU.

\section*{Declarations}
The author has no conflicts of interest to declare. Data sharing is not applicable
to this article as no new datasets were generated or analyzed during the
current study.

\bigskip

\bibliography{Review}

\end{document}